\documentclass[11pt]{article}
\usepackage[dvips]{graphicx}
\usepackage[english]{babel}

\usepackage{epsfig}
\usepackage{graphicx}
\usepackage{indentfirst}
\usepackage{amsmath}
\usepackage{amsfonts}
\usepackage{amssymb}

\begin{document}
\begin{center}
{\Large\bf Finite-time singularities in f(R, T) gravity and the effect of conformal anomaly\\}
\medskip

M. J. S. Houndjo ${}^{a,b,}$\footnote{e-mail:
sthoundjo@yahoo.fr},\, C. E. M. Batista $^{ c,}$\footnote{e-mail:
cedumagalhaes22@hotmail.com}, J. P. Campos $^{ d,}$\footnote{e-mail: jpcampospt@gmail.com} and O. F. Piattella $^{ e,}$\footnote{e-mail: oliver.piattella@ufes.br}\\ 
$^a$ { \it Departamento de Engenharia e Ci\^{e}ncias Naturais - CEUNES \\
Universidade Federal do Esp\'irito Santo\\
CEP 29933-415 - S\~ao Mateus - ES, Brazil}\\
$^b$ {\it Institut de Math\'{e}matiques et de Sciences Physiques (IMSP)}\\
 {\it 01 BP 613 Porto-Novo, B\'{e}nin}
\\
$^c$ {\it Departamento de F\'{i}sica, Universidade Estadual de Feira de Santana, BA, Brazil}\\

$^d$ {\it Centro de Ci\^encias Exatas e Tecnol\'ogicas, Universidade Federal do Rec\^oncavo da Bahia, Cruz das Almas, BA, Brazil}\\  
$^e$ {\it Departamento de F\'{i}sica, Universidade Federal do Esp\'irito Santo, Vit\'oria, ES, Brazil}\\
\date{}

\end{center}
\begin{abstract}
We investigate $f(R,T)$ gravity models (where $R$ is the curvature scalar and $T$ is the trace of the stress-energy tensor of ordinary matter) that are able to reproduce the four known types of future finite-time singularities. We choose a suitable expression for the Hubble parameter in order to realise the cosmic acceleration and we introduce two parameters, $\alpha$ and $H_s$, which characterise each type of singularity. We address conformal anomaly and we observe that it cannot remove the Sudden Singularity or the Big Brake, but, for some values of $\alpha$, the Big Rip and the Big Freeze may be avoided. We also find that, even without taking into account conformal anomaly, the Big Rip and the Big Freeze may be removed thanks to the presence of the $T$ contribution of the $f(R,T)$ theory.
\end{abstract}

Pacs numbers: 04.50.Kd, 95.35.+d, 95.36.+x, 98.80.Qc

\section{Introduction}

Observation indicates that the current expansion of the universe is accelerating \cite{bamba1, bamba2}. There are two large branches of cosmology attempting to explain this phenomenon \cite{bamba3}-\cite{bamba12}: $i)$ Introducing dark energy in the framework of general relativity; $ii)$ investigating modified versions of the gravitational theory. In this paper we follow the latter approach. The most widely renown modified theory is $f(R)$-gravity, in which the action is determined by an arbitrary function $f(R)$ of the Ricci scalar $R$ \cite{bamba5}-\cite{bamba8}. Another interesting modified theory of gravity is $f(G)$-gravity, where $G = R^2-4R_{\mu\nu}R^{\mu\nu} + R_{\mu\nu\rho\sigma}R^{\mu\nu\rho\sigma}$ is the Gauss-Bonnet invariant ($R_{\mu\nu}$ and $R_{\mu\nu\rho\sigma}$ are the Ricci tensor and the Riemann tensor respectively).\par
In the framework of general relativity, it is well known that a phantom phase usually ends up in the type I (Big Rip) finite-time future singularity (FTFS) \cite{bamba16, stephane1}. It has also been demonstrated, e.g. in \cite{bamba17}-\cite{bamba19}, that the dark energy equation of state parameter may show all four possible types of FTFS and that there is no qualitative difference between dark energy and modified gravity. Thus, $f(R)$-gravity dark energy models also may bring the universe evolution to all four possible FTFS \cite{bamba20}-\cite{bamba24}. Another interesting aspect of $f(R)$ modified gravity is that it may provide models that remove FTFS by adding, for example, a $R^2$-term \cite{bamba20, bamba21, bamba23, bamba25}. { \bf Still in the framework of $f(R)$ gravity, different aspects from cosmic strings to symmetry have been investigated \cite{davood} }. The $f(G)$-gravity also is a class of modified theories that may lead to finite-time future singularities \cite{bamba21, bamba0}. {\bf A new type of modified gravity has now received attention, that is the so-called $f(\mathcal{T})$ gravity and considerable results have been found \cite{momenisuggestions}, where $\mathcal{T}$ is the torsion scalar. On the other hand, some types of singularities have been investigated in the framework of Ho$\check {\textbf{r}}$ava-Lifshitz gravity and interesting results have been obtained \cite{momenisuggestions2}. }\par

In the present paper, we approach $f(R, T)$-gravity, where $T$ is the trace of the energy-momentum tensor of ordinary matter, and investigate models that may lead to the four FTFS. This type of modified theory of gravity has been first developed by Harko \textit{et al.} \cite{harko}, who derive the gravitational field equations in the metric formalism, as well as the equations of motion of test particles, which follow from the covariant divergence of the stress-energy tensor. They also analysed the Newtonian limit of the equations of motion, and provide a constraint on the magnitude of the extra-acceleration by investigating the perihelion precession of Mercury. In \cite{stephane7}, $f(R, T)$ models have been constructed describing the transition from the matter-dominated phase to the late-times accelerated one and in \cite{oliver} the authors consider cosmological scenarios based on $f(R, T)$ theory and the function $f(R, T)$ is numerically reconstructed from holographic dark energy. {\bf On the other hand, Jamil et al shown that $f(R,T)$ gravity violates the first law of thermodynamics but it reproduces all cosmological epochs \cite{momenilaw}. }\par
In this paper, we consider the special case in which the function $f(R, T)$ is the usual Einstein-Hilbert term plus a correction $g(T)$, i.e. $f(R, T) = R + 2g(T)$. A suitable expression is assumed for the Hubble parameter, which may provide the four FTFS. For some values of the parameter $\alpha$, differential equations of $g(T)$ are established and solved. Another aspect which we study are quantum effects due to conformal anomaly near the singularities. We observe that conformal anomaly cannot remove the sudden singularity or the type IV singularity (Big Brake). However, for some values of the parameter $\alpha$, the Big Rip and the type III singularity (Big Freeze) may be avoided. Another important result that we obtain is that, even without taking into account quantum effects, the Big Rip and the Big Freeze may be avoided thanks to the contribution $T$ of the modified theory.\par
The paper is organized as follows. In Sec. $2$, we present the general formulation of the theory addressing the special case $f(R, T)= R+2g(T)$ and obtaining the models that reproduce each type of FTFS. In Sec. $3$, we investigate quantum effects due to conformal anomaly. We present our conclusions and perspectives in Sec. $4$.

\section{f(R, T)-gravity}

We assume a modification of general relativity in which the Ricci scalar $R$ is replaced by an arbitrary function $f(R, T)$. Then, the usual Einstein-Hilbert action turns to
\begin{eqnarray}\label{edu1}
S=\frac{1}{2}\int d^4x \sqrt{-g}\left[f(R, T)+\mathcal{L}_m\right]\,,
\end{eqnarray}
where $\mathcal{L}_m$ is the matter Lagrangian density and $T=g^{\mu\nu}T_{\mu\nu}$ is the trace of the matter energy-momentum tensor $T_{\mu\nu}$ which is defined as 
\begin{eqnarray}\label{edu2}
T_{\mu\nu}=-\frac{2}{\sqrt{-g}}\frac{\delta(\sqrt{-g}\mathcal{L}_m)}{\delta g^{\mu\nu}}\,\,.
\end{eqnarray}
Varying the action (\ref{edu1}) with respect to the metric, one obtains the gravitational field equations
\begin{eqnarray}\label{edu3}
f_R(R, T)R_{\mu\nu}-\frac{1}{2}f(R, T)g_{\mu\nu}+\left( g_{\mu\nu}\Box -\nabla_\mu\nabla_\nu\right) f_R(R, T) = \nonumber\\ T_{\mu\nu}-f_T(R, T)T_{\mu\nu}-f_T(R, T)\Theta_{\mu\nu}\,\,,
\end{eqnarray}
where $f_R$ and $f_T$ denote the derivatives of $f$ with respect to $R$ and $T$, respectively; $\nabla_{\mu}$ is the covariant derivative and $\Theta_{\mu\nu}$ is defined by 
\begin{eqnarray}\label{edu4}
\Theta_{\mu\nu}\equiv g^{\alpha\beta}\frac{\delta T_{\alpha\beta}}{\delta g^{\mu\nu}}= -2T_{\mu\nu}+g_{\mu\nu}\mathcal{L}_m-2g^{\alpha\beta}\frac{\partial^{2}\mathcal{L}_m}{\partial g^{\mu\nu}\partial g^{\alpha\beta}}\,\,.
\end{eqnarray}
We adopt the flat Friedmann-Lema\^itre-Robertson-Walker (FLRW) metric 
\begin{equation}
 ds^2 = dt^2 - a^2(t)\delta_{ij}dx^idx^j\;,
\end{equation}
as the space-time geometry, and we assume the matter content of the universe to be a perfect fluid, whose stress-energy tensor has the following form:
\begin{eqnarray}\label{edu5}
T_{\mu\nu}=(\rho+p)u_\mu u_\nu-pg_{\mu\nu}\,\,,
\end{eqnarray}
where $u_{\mu}$ is the four-velocity and $\rho$ and $p$ are the energy density and the pressure of ordinary matter, respectively. The matter Lagrangian density can be taken as $\mathcal{L}_m=-p$, which implies $\Theta_{\mu\nu}=-2T_{\mu\nu}-pg_{\mu\nu}$. Thus, Eq.~(\ref{edu3}) becomes
\begin{eqnarray}\label{edu6}
f_R(R, T)R_{\mu\nu}-\frac{1}{2}f(R, T)g_{\mu\nu}+\left( g_{\mu\nu}\Box -\nabla_\mu\nabla_\nu\right) f_R(R, T) =\nonumber\\ T_{\mu\nu}+f_T(R, T)T_{\mu\nu}+pg_{\mu\nu}f_T(R, T)\,.
\end{eqnarray}
Our ansatz for the function $f$ is the following: $f(R, T)= R + 2g(T)$, where $g(T)$ is an arbitrary function of $T$. Then, Eq.~(\ref{edu6}) becomes
\begin{eqnarray}\label{edu7}
R_{\mu\nu}-\frac{1}{2}Rg_{\mu\nu}=T_{\mu\nu}+2g_T(T)T_{\mu\nu}
+\left[2pg_T(T)+g(T)\right]g_{\mu\nu}\,.
\end{eqnarray}
The components $00$ and $ii$ of Eq.~(\ref{edu7}) read
\begin{eqnarray}
3H^2&=&\rho_{eff} = \rho+2\left(\rho+p\right)g_T(T)+g(T)\;,\label{edu8}\\
-2\dot{H}-3H^2&=&p_{eff} = p-g(T)\label{edu9}\,\,,
\end{eqnarray}
where we have introduced $\rho_{eff}$ and $p_{eff}$ as the effective energy density and pressure, respectively. Combining Eqs.~(\ref{edu8}) and (\ref{edu9}), we get the following equations 
\begin{eqnarray}
-2\dot{H}=\rho\left(1+\omega\right)\left[1+2g_T(T)\right]\;,
\label{edu10}\\
6\frac{\ddot{a}}{a}=2\left[\rho+g(T)\right]-\rho\left(1+\omega\right)\left[2g_T(T)+3\right]\,,\label{edu11}
\end{eqnarray}
where we have defined $\omega \equiv p/\rho$.\par
Note that, in order to produce an accelerated expansion, the function $g(T)$ has to satisfy the condition
\begin{eqnarray}\label{edu12}
2g(T)-2\rho\left(1+\omega\right)g_T(T)>\rho\left(1+3\omega\right)\;.
\end{eqnarray}

\section{Finite-time future singularities in $R + 2g(T)$ gravity}

We are interested in the modified $f(R, T)$ gravity models that produce some types of finite-time future singularities (see \cite{gorbu6,gorbu7}), given by the Hubble parameter
\begin{eqnarray}
H=h(t_s-t)^{-\alpha}\,\,\,,\label{edu13}
\end{eqnarray}
where $h$ and $t_s$ are positive constants and  $t < t_s$. This expression is chosen for guaranteeing an expanding universe. Due to fact that $H$ or some of its derivatives (and therefore the curvature) could become singular when $t$ is close to $t_s$, the parameter $\alpha$ could be either positive or negative. If $\alpha > 0$, then $H$ diverges as the singularity time is approached. The same expression of the Hubble parameter provides the acceleration of the universe, since the strong energy condition is violated. On the other hand, when $\alpha < 0$, a relevant positive constant $H_s$ may be introduced and interpreted as the Hubble parameter at the singularity time, such that $H = h(t_s-t)^{-\alpha} + H_s$. In this case, some derivatives of $H$ become singular. Note that the case $\alpha=0$ corresponds to the de Sitter space (which we do not treat in this paper).\par 
Depending on the value of $\alpha$, different expressions can be found for the scale factor. \par
If $\alpha=1$, the scale factor reads 
\begin{eqnarray}
a(t)= \bar{a}(t_s-t)^{-h}\,\,\,,\label{edu14}
\end{eqnarray}
where $\bar{a}$ is an integration constant.\par
If $\alpha>0$ and $\alpha\neq 1$,  we get 
\begin{eqnarray}\label{edu15}
a(t)= \bar{a} \exp\left[\frac{h(t_s-t)^{1-\alpha}}{\alpha-1}\right]\,\,\,.
\end{eqnarray}
Finally, if $\alpha<0$ and $H_s>0$, the scale factor behaves as 
\begin{eqnarray}\label{edu16}
a(t) = \bar{a}\exp\left\{-(t_s-t)\left[H_s-\frac{h(t_s-t)^{-\alpha}}{\alpha-1}\right]\right\}\,\,\,.
\end{eqnarray}
The finite-time future singularities can be classified as follows \cite{gorbu8, steph4}:\par
$\bullet$ Type I (Big Rip): for $t\rightarrow t_s$, $a\rightarrow\infty$, $\rho_{eff}$ and $|p_{eff}|\rightarrow \infty$ at $t=t_s$. This corresponds to $\alpha \ge 1$.\par
$\bullet$ Type II (Sudden): for $t\rightarrow t_s$, $a\rightarrow a_s$, $\rho_{eff}\rightarrow \rho_s$ and $|p_{eff}|\rightarrow \infty$. It corresponds to $-1<\alpha< 0$.\par
$\bullet$ Type III (Big Freeze): for $t\rightarrow t_s$,  $a\rightarrow a_s$,  $\rho_{eff}\rightarrow \infty$ and $|p_{eff}|\rightarrow \infty$. This corresponds to $0<\alpha<1$. \par
$\bullet$ Type IV (Big Brake): for $t\rightarrow t_s$, $a\rightarrow a_s$, $\rho_{eff}\rightarrow 0$, $p_{eff}\rightarrow 0$ and higher derivatives of $H$ diverge. This corresponds to the case $\alpha<-1$, but with $\alpha$ not integer.\par
We investigate $f(R,T) = R + 2g(T)$ gravity models which generate the above types of finite-time singularities. This corresponds to determine which form of $g(T)$ gives rise to each type of singularity.\par

\subsection{Treating the case $a(t)= \bar{a}(t_s-t)^{-h}$}

With this scale factor, the energy density and the first derivative of the Hubble parameter can be written as
\begin{eqnarray}
\rho(t)&=&\rho_0\bar{a}^{-3(1+\omega)}\left(t_s-t\right)^{3h(1+\omega)}\,\,\,.\label{edu17}\\
\dot{H}&=&  h\left(t_s-t\right)^{-2}\,\,\,.\label{edu18}
\end{eqnarray}
Substituting Eqs.~(\ref{edu17}) and (\ref{edu18}) into Eq.~(\ref{edu10}), one obtains
\begin{eqnarray}
2g_T(T)+\frac{2 h \bar{a}^{3(1+w)}}{\rho_0(1+w)}\left(t_s-t\right)^{-2-3h(1+w)}+1=0\,\,\,. \label{edu19}
\end{eqnarray}
On the other hand, we can write the trace of the energy-momentum tensor in terms of the energy density as
\begin{equation}\label{edu20}
T = \rho-3p = \rho\left(1-3\omega\right)\,\,\,,
\end{equation}
from which one has $\rho=T/(1-3\omega)$, and then
\begin{eqnarray}\label{edu21}
t_s-t= \bar{a}^{\frac{1}{h}}\left[\rho_0(1-3\omega)\right]^{-\frac{1}{3h(1+\omega)}}T^{\frac{1}{3h(1+\omega)}}\,\,\,.
\end{eqnarray}
By injecting Eq.~(\ref{edu21}) into Eq.~(\ref{edu19}), we get 
\begin{equation}\label{edu22}
 2g_T(T)+\lambda T^{\sigma}+1 = 0\,\,\,,
\end{equation}
where
\begin{equation}
\lambda = \frac{2 h(1-3\omega) \bar{a}^{-\frac{2}{h}}}{1+\omega}\left[\rho_0(1-3\omega)\right]^{\frac{2}{3h(1+\omega)}}\,\,,\quad \sigma=-\frac{2}{3h(1+\omega)}-1\,\,\,, 
\end{equation}
and whose general solution reads
\begin{eqnarray}
g(T)=\frac{1}{2}\left(-T-\frac{\lambda T^{1+\sigma}}{1+\sigma}\right)+C_1\,\,\,,\label{edu23}
\end{eqnarray}
and the corresponding $f(R,T)$ model reads
\begin{eqnarray}\label{edu24}
f(R,T) = R - T-\frac{\lambda T^{1+\sigma}}{1+\sigma}+2C_1\,\,\,,
\end{eqnarray}
where $C_1$ is an integration constant. Let us assume that for the present time $t_0$, the corresponding value of the trace is $T_0$. When $g(T) = 0$, the Einstein theory is recovered and at $t = t_0$ Eq.~(\ref{edu8})reads $3H^2_0=\rho_0$ (we normalise the present time scale factor to unity). In the general case, working with the model $R+2g(T)$, the initial condition has to be the same as in Einstein theory, that is, one has to get $3H^2_0=\rho_{eff\,0}$. This leads to 
\begin{eqnarray}\label{edu25}
2\rho_0\left(1+\omega\right)g_T(T_0)+g(T_0)=0\,\,\,.
\end{eqnarray}
By making use of (\ref{edu23}), one acquires
\begin{eqnarray}
C_1=\frac{3(3-\omega)}{2}H_{0}^{2}+h\bar{a}^{-\frac{2}{h}}\left[2-3h(1-3\omega)\right]\,\,\,.\label{edu26}
\end{eqnarray}
Here, $T_0$ may be determined in terms of $\rho_0$, which in turn may be calculated from the current value of the Hubble parameter $H_0=2.1h\times 10^{-42}$GeV \cite{51thermo2}, with $h=0.7$ in \cite{secondde3thermo2} through the relation $3H^2_0=\rho_0$. Then the model (\ref{edu24}), with $C_1$ given by (\ref{edu26}), can lead to the Big Rip.
\subsection{Treating the case $a(t)= \bar{a}\exp\left[\frac{h(t_s-t)^{1-\alpha}}{\alpha-1}\right]$}
In this case, the appearance of both the Big Rip ($\alpha>1$) and the Big Freeze  ($0<\alpha<1$) is possible. The energy density and the first derivative of the Hubble parameter read
\begin{eqnarray}\label{edu27}
\rho&=&\rho_0\bar{a}^{-3(1+w)}\exp\left[\frac{-3h(1+w)(t_s-t)^{1-\alpha}}{\alpha-1}\right]\label{edu27}\;,\\
\dot{H}&=&\alpha h\left(t_s-t\right)^{-\alpha-1}\,\,\,. \label{edu28}
\end{eqnarray}
By making use of Eqs.~(\ref{edu27}) and (\ref{edu28}), Eq.~(\ref{edu10}) becomes
\begin{eqnarray}
2g_T(T)+\frac{2\alpha h\bar{a}^{3(1+\omega)}(t_s-t)}{\rho_0(1+\omega)}\exp\left[\frac{3h(1+\omega)(t_s-t)^{1-\alpha}}{\alpha-1}\right]+1=0\,\,\,. \label{edu29}
\end{eqnarray}
Using the relation $\rho=T/(1-3w)$, and Eq.~(\ref{edu27}), one gets
\begin{eqnarray}
t_s-t=\ln^{\frac{1}{1-\alpha}}{\left[[\rho_0(1-3\omega)]^{\frac{\alpha-1}{3h(1+\omega)}}\bar{a}^{\frac{1-\alpha}{h}}\,T^{\frac{1-\alpha}{3h(1+\omega)}}\right]}\,\,\,.\label{edu30}
\end{eqnarray}
Substituting Eq.~(\ref{edu30}) into Eq.~(\ref{edu29}), we acquire
\begin{eqnarray}
2g_T(T)+\frac{2\alpha h(1-3\omega)}{(1+\omega)T}\ln^{\frac{1}{1-\alpha}}{\left[[\rho_0(1-3\omega)]^{\frac{\alpha-1}{3h(1+\omega)}}\bar{a}^{\frac{1-\alpha}{h}}\,T^{\frac{1-\alpha}{3h(1+\omega)}}\right]}+1=0\,\,\,\,,\label{edu31}
\end{eqnarray}
whose general solution reads
\begin{eqnarray}
g(T)=-\frac{T}{2}-\frac{3\alpha h^2(1-3\omega)}{2-\alpha}\ln^{\frac{2-\alpha}{1-\alpha}}{\left[[\rho_0(1-3\omega)]^{\frac{\alpha-1}{3h(1+\omega)}}\bar{a}^{\frac{1-\alpha}{h}}\,T^{\frac{1-\alpha}{3h(1+\omega)}}\right]}+C_2\,\,\,,\label{edu33}
\end{eqnarray}
where $C_2$ is an integration constant. Thus the corresponding $f(R,T)$ model is
\begin{eqnarray}
f(R,T)=R-T-\frac{6\alpha h^2(1-3\omega)}{2-\alpha}\ln^{\frac{2-\alpha}{1-\alpha}}{\left[[\rho_0(1-3\omega)]^{\frac{\alpha-1}{3h(1+\omega)}}\bar{a}^{\frac{1-\alpha}{h}}\,T^{\frac{1-\alpha}{3h(1+\omega)}}\right]}+2C_2\,\,\,,\label{edu34}
\end{eqnarray}
By using the same initial condition  as made in (\ref{edu25}), one obtains
\begin{eqnarray}
C_2=\frac{3(3-\omega)}{2}H^2_0+2\alpha h\ln^{\frac{1}{1-\alpha}}{\left(\bar{a}^{\frac{1-\alpha}{h}}\right)}\left[ \frac{3h(1-3\omega)}{2(2-\alpha)}\ln{\left(\bar{a}^{\frac{1-\alpha}{h}}\right)-1}\right] \,\,\,\,.\label{edu34}
\end{eqnarray}
We conclude that the model (\ref{edu33}) can allow the occurrence of  both the Big Rip ($\alpha>1$) and the Big Freeze ($0<\alpha<1$).

\subsection{Treating the case\quad $a(t)=\bar{a}\exp\left\{-(t_s-t)\left[H_s-\frac{h(t_s-t)^{-\alpha}}{\alpha-1}\right]\right\}$}
Here the first derivative of the Hubble parameter remains the same as in (\ref{edu28}), but the energy density is written as
\begin{eqnarray}
\rho=\rho_0\bar{a}^{-3(1+\omega)}\exp\left\{3(1+\omega)(t_s-t)\left[H_s-\frac{h(t_s-t)^{-\alpha}}{\alpha-1}\right]\right\} \,\,\,. \label{edu35}
\end{eqnarray}
Here, note that it is not easy to express $(t_s-t)$ in terms  of $\rho$ or $T$, due to the complicated form of (\ref{edu35}). However, one can try to face the problem in other angle, i.e. finding the algebraic function $R+2g(T)$ around the singularities (Sudden and Big Brake). Since $\alpha<0$, one always gets  $1-\alpha >1$. Then near the singularity, (\ref{edu35}) reduces to 
\begin{eqnarray}
\rho=\rho_0\bar{a}^{-3(1+\omega)}\exp\left\{3H_s(1+\omega)(t_s-t)\right\}\,\,\,\label{edu36}
\end{eqnarray}
from which, one extracts 
\begin{eqnarray}
t_s-t=\ln{\left[[\rho_0(1-3\omega)]^{-\frac{1}{3H_s(1+\omega)}}\bar{a}^{\frac{1}{H_s}}T^{\frac{1}{3H_s(1+\omega)}}\right]}\,\,\,.\label{edu37}
\end{eqnarray}
By using (\ref{edu37}), (\ref{edu28}) becomes
\begin{eqnarray}
\dot{H}=\alpha h \ln^{-\alpha-1}{\left[[\rho_0(1-3\omega)]^{-\frac{1}{3H_s(1+\omega)}}\bar{a}^{\frac{1}{H_s}}T^{\frac{1}{3H_s(1+\omega)}}\right]}\,\,\,,\label{edu38}
\end{eqnarray}
and the differential equation (\ref{edu10}) reads
\begin{eqnarray}
g_T(T)+\frac{2\alpha h(1-3\omega)}{(1+\omega)T}\ln^{-\alpha-1}{\left[[\rho_0(1-3\omega)]^{-\frac{1}{3H_s(1+\omega)}}\bar{a}^{\frac{1}{H_s}}T^{\frac{1}{3H_s(1+\omega)}}\right]}+1=0\,\,\,.\label{edu39}
\end{eqnarray}
The general solution of (\ref{edu39}) is 
\begin{eqnarray}
g(T)=-\frac{T}{2}+3hH_s(1-3\omega)\ln^{-\alpha}{\left[[\rho_0(1-3\omega)]^{-\frac{1}{3H_s(1+\omega)}}\bar{a}^{\frac{1}{H_s}}T^{\frac{1}{3H_s(1+\omega)}}\right]}+C_3\,\,\,,\label{edu40}
\end{eqnarray}
where $C_3$ is an integration constant. The $f(R,T)$ corresponding model reads
\begin{eqnarray}
f(R,T)=R-T+6hH_s(1-3\omega)\ln^{-\alpha}{\left[[\rho_0(1-3\omega)]^{-\frac{1}{3H_s(1+\omega)}}\bar{a}^{\frac{1}{H_s}}T^{\frac{1}{3H_s(1+\omega)}}\right]}+2C_3\,\,\,.\label{edu41}
\end{eqnarray}
As for the previous cases, the constant $C_3$ is determined from the initial condition (\ref{edu25}), as
\begin{eqnarray}
C_3=-\frac{3(7\omega+3)}{2}H^2_0+h\ln^{-\alpha-1}{\left(\bar{a}^{\frac{1}{H_s}}\right)}\Big\{-4\alpha
-3H_s(1-3\omega)\ln{\left(\bar{a}^{\frac{1}{H_s}}\right)}\Big\}\,\,\,.\label{edu42}
\end{eqnarray}
Hence, the model (\ref{edu42}) can allow the appearance of both the Sudden singularity (for $-1<\alpha<0$) and the Big Brake ($\alpha <-1$).


\section{Quantum effects near finite time singularity}

We may check the avoidance of finite time future singularities by taking into account quantum effects. In this work we address conformal anomaly. Near the future finite-time singularity ($t\rightarrow t_s$) the curvature diverges. Due to their dependence on the curvature, quantum effects coming from conformal anomaly also become important. In such a situation, all classical considerations have to be revised and any claim about the appearance of future finite-time singularity cannot be justified without an account of quantum effects. One may incorporate the massless quantum effects by taking into account the conformal anomaly contribution as a backreaction near the singularity. 
{\bf The viability and importance of these terms in now era, for example about the effects of conformal anomaly  on black hole have widely investigated in \cite{momenica}.}

The conformal anomaly $T_A$ has the following well-known expression \cite{conformal20}
\begin{eqnarray}\label{edu43}
T_A=b\left(F+\frac{2}{3}\Box R\right)+b^{\prime}G+b^{\prime\prime}\Box R\,\,\,\,,
\end{eqnarray}
where $F$ is the square of the four-dimensional Weyl tensor and $G$ is Gauss-Bonnet invariant, given by
\begin{eqnarray}
F&=&\frac{1}{3}R^2-2R_{\mu\nu}R^{\mu\nu}+R_{\mu\nu\rho\sigma}R^{\mu\nu\rho\sigma}\,\,\,,\label{edu44}\\
G&=&R^2-4R_{\mu\nu}R^{\mu\nu}+R_{\mu\nu\rho\sigma}R^{\mu\nu\rho\sigma}\,\,\,.\label{edu45}
\end{eqnarray}
Explicitly, if there are $N$ scalars, $N_{1/2}$ spinors, $N_1$ vector fields, $N_2$ gravitons and $N_{HD}$ higher derivative conformal scalars, $b$ and $b^{\prime}$ are expressed as 
\begin{eqnarray}
b&=&\frac{N+6N_{1/2}+12N_1+611N_2-8N_{HD}}{1920\pi^2}\;,\label{edu46}\\
b^{\prime}&=&\frac{N+11N_{1/2}+62N_1+1411N_2-28N_{HD}}{5760\pi^2}\,\,\,,\label{edu47}
\end{eqnarray}
whereas $b^{\prime\prime}$ is an arbitrary constant whose value can be shifted by the finite renormalization of the local counter-term  $R^2$.\par
Now, by taking into account quantum contribution from the conformal anomaly, Eq.~(\ref{edu8}) is modified and its trace can be written as
\begin{eqnarray}\label{edu69}
-R= \left[1+2g_T(T)\right]T+8pg_T(T)+4g(T)+T_A\,\,\,\,.\label{edu48}
\end{eqnarray}
By using again the barotropic state equation $p=\omega \rho$ for the ordinary matter content, one acquires
\begin{eqnarray}
-R=T+\left[\frac{2(1+\omega)}{1-3\omega}\right]Tg_T+4g+T_A\,\,\,\,.
\end{eqnarray}
For the FLRW universe, it is clear that 
\begin{eqnarray}
F=0\,\,, \quad G=24\left(\dot{H}H^2+H^4\right)\,\,\,.
\end{eqnarray}
Let us focus our attention on the case $2b+3b^{\prime\prime}=0$ and set $b^{\prime}=1$. Then, one gets $T_A=G$. We may analyse quantum effects coming from conformal anomaly near each type of singularity. Two important cases need to be considered in doing this analysis, $\omega<-1$ and $\omega>-1$. \par

\subsection{The case $\omega<-1$}

\subsubsection{Conformal anomaly near the Big Rip}

In this case, one can observe two conditions on the parameter $\alpha$, that is, $\alpha=1$ and $\alpha>1$.\par
$\bullet$ The case $\alpha=1$\par
Here, the curvature behaves as $R\propto (t_s-t)^{-2}$, the trace as $T\propto (t_s-t)^{3h(1+\omega)}$, and $T_A\propto (t_s-t)^{-4}$. \par
Here, we may distinguish three conditions; $-2<3h(1+\omega)<0$, $3h(1+\omega)=-2$ and $3h(1+\omega)<-2$.\par
For $-2<3h(1+\omega)<0$, one has $g(T)\propto Tg_T(T)\propto (t_s-t)^{-2}$. It appears that, as the singularity time is approached, the conformal anomaly term dominates over other terms. Hence, the Big Rip may be avoided.\par
For $3h(1+\omega)=-2$,  one gets $g(T)\propto Tg_T(T) \propto (t_s-t)^{-2}$. Also, in this case, the conformal anomaly term dominates and the Big Rip may also be avoided. \par 
For $3h(1+\omega)<-2$, the corresponding expression of $g(T)$ is that coming from Eq.~(\ref{edu38}), and $g(T)\propto Tg_T(T)\propto (t_s-t)^{3h(1+\omega)}$. Here, it can be observed that, even if the conformal anomaly is not taken into account the Big Rip may be avoided from the contribution of the matter in the gravitational part of the action. This can be viewed as a reason for taking into account the trace $T$ term in the gravitational part since, in this case, the Big Rip can be avoided without the need of quantum effect.\par
$\bullet$ The case $\alpha>1$\par
Here, as the singularity time is approached, the curvature behaves as $R\propto (t_s-t)^{-2\alpha}$, the conformal anomaly as $T_A\propto (t_s-t)^{-4\alpha}$, and the ordinary trace as $T \propto \exp{\left\{\frac{-3h(1+\omega)}{\alpha-1}(t_s-t)^{1-\alpha}\right\}}$. We also obtain  $g(T)\propto Tg_T(T)\propto \exp{\left\{\frac{-3h(1+\omega)}{\alpha-1}(t_s-t)^{1-\alpha}\right\}}$, that one can assimilate to $(t_s-t)^{-2N}$, where $N=N(\alpha)$ is a positive function of $\alpha$. This assumption is possible since, for $\alpha>1$, $1-\alpha<0$, and also, as we are dealing with the case $\omega<-1$, one gets $-3h(1+\omega)>0$. On the other hand, it is clear here that we also have $-4\alpha< -2\alpha$, and then, the conformal anomaly term dominates over the curvature one. One can conclude that for $\alpha>1$, quantum effects from conformal anomaly may allow the avoidance of the Big Rip. Note also that, if for some values of $\alpha>1$ on acquires $N>2$, then, we regain the situation in which $f(R,T)$ itself lead to the avoidance of the Big Rip, thanks to the ordinary trace terms in the gravitational part of the action, without the need of quantum effects.

\subsubsection{Conformal anomaly near the sudden singularity}

In this case, $-1<\alpha<0$, and as the singularity is approached, the curvature behaves as $R \propto (t_s-t)^{-\alpha-1}$, the ordinary trace behaves as a constant, and the conformal anomaly as $T_A \propto (t_s-t)^{-3\alpha-1}$.  Here, one gets $g_T\propto (t_s-t)^{-\alpha-1}$ and $ g\propto (t_s-t)^{-\alpha}$. Note that for $-1<\alpha<0$, we always have $-3\alpha-1> -\alpha-1$. This means that the sudden singularity cannot be avoided from conformal anomaly contribution. This results perfectly agrees with that found in \cite{steph4,barrow1}, where in the framework of GR it has been shown that sudden singularity cannot be avoided both from particle production and conformal anomaly. However, it appears that the avoidance of the sudden singularity could come from the matter contribution, due to the same exponent for $R$ and $g_T$.

\subsubsection{Conformal anomaly near the Big Freeze}

This type of singularity occurs for $0<\alpha<1$, and around the singularity, the curvature behaves as $R\propto (t_s-t)^{-\alpha-1}$, the conformal anomaly as $T_A\propto (t_s-t)^{-3\alpha-1}$. Here, the ordinary trace behaves as constant and, $g_T\propto (t_s-t)$ and $g\propto (t_s-t)^{2-\alpha}$. As $0<\alpha<1$, one gets $-3\alpha-1<-\alpha-1<0$. Then, the Big Freeze  may be avoided from conformal anomaly effects. 

\subsubsection{Conformal anomaly near the Big Brake}

The condition of occurrence of this type of singularity is $\alpha<-1$, and in this case the curvature $R$, the conformal anomaly $T_A$, $g_T$ and $g$ are all non-singular, while  from the first to higher of $R$ as the singularity time is approached. Hence the singularity is always robust against the conformal anomaly and cannot be avoided.\par

\subsection{Commenting the case $\omega>-1$}

In this case, $3h(1+\omega)$ is always positive. This may certainly change some of the results obtained in the previously case. We also analyse here the conformal anomaly effect near each singularity.
\subsubsection{Conformal anomaly around the Big Rip}
$\bullet$ The case $\alpha=1$\par
Here, as the singularity is approached, the curvature scalar behaves as $R\propto (t_s-t)^{-2}$, the ordinary trace is finite, the conformal anomaly behaves as $T_A=(t_s-t)^{-4}$, $Tg_T\propto g\propto (t_s-t)^{-\frac{2}{3h(1+\omega)}}$. In this case, two conditions my be distinguished, the first when $0<3h(1+\omega)<1$ and the second when $3h(1+\omega)>1$.\par
For the case where $0<3h(1+\omega)<1$,  the term $(t_s-t)^{-\frac{2}{3h(1+\omega)}}$ diverges more than the curvature as the singularity is approached. Hence, not only the conformal anomaly may prevent the Big Rip, but also this singularity may be avoided from the $T$ terms contribution, even quantum effects are nor taken into account.\par
For $3h(1+\omega)>1$, just the conformal anomaly term dominates over the curvature term, and may prevent the singularity.\par
$\bullet$ The case $\alpha>1$\par
Here, as the singularity time is approached, the curvature behaves as $R\propto (t_s-t)^{-2\alpha}$, the conformal anomaly as $T_A\propto (t_s-t)^{-4\alpha}$, and the ordinary trace and the product $Tg_T$ are finite, while $g(T)\propto (t_s-t)^{2-\alpha}$. Note that for any $\alpha>1$, one always gets $2-\alpha>-2\alpha$. Then, the curvature term always dominates over the $T$ contribution terms, while it diverges less than the conformal anomaly term. Hence, the only conformal anomaly effects can present the Big Rip.

\subsubsection{Conformal anomaly near the Sudden singularity}
In this case, since the term $3h(1+\omega)$ is not present, we regain the same situation as in the case where $\omega<-1$.

\subsubsection{Conformal anomaly around the Big Freeze}
Here the situation is the same as for that of the case $\omega<-1$. Then the Big Freeze only may be avoided from the conformal anomaly effects.

\subsubsection{Conformal anomaly near the Big Brake}
Also here, since the term $3h(1+\omega)$ is not present, the same situation as in the case where $\omega<-1$ occurs.

\section{Conclusion}
We considered the modified $f(R, T)$ theory of gravity where $R$ is the curvature scalar and $T$ the trace of the energy-momentum tensor. We focused our attention in reconstructing the models of this theory, that may lead to finite-time future singularities. We assumed a suitable expression for the Hubble parameter in order to provide the expansion of the universe and realise the four finite-time future singularities, by adjusting an input parameter $\alpha$. A special expression $f(R, T)= R+2g(T)$ is assumed, viewed as a $T$-term correction to the Einstein-Hilbert term, $R$. From this, differential equation for $g(T)$ are obtained through the equations of motion, for some values of the parameter $\alpha$. In each case, the differential equation is solved leading to the $f(R, T)$ model that leads to finite-time future singularity. \par
As the curvature diverges at singularity time, quantum effect may be introduced. In this way, conformal anomaly is introduced and its effects at singularity time in analysed. We observed that conformal anomaly cannot remove the sudden singularity nor the Big Brake. However, for some values of the parameter $\alpha$, the Big Rip and the Big Freeze may be avoided. Another important result that we find is that, even without the contribution of quantum effect, the Big Rip and the Big Freeze may be avoided from the contribution of the ordinary trace terms, for some values of the parameter $\alpha$. This is a reason for considering $f(R, T)$ in the study of some scenarios of evolution of the universe.\par
In the case of Sudden singularity and the Big Brake, we observed that neither conformal anomaly or the terms coming from the classical trace allow the avoidance of these singularities. Still, in order to  check the possible avoidance of the Sudden and the Big Brake, the viscosity of the fluid characterising the matter content may be introduced. We leave this as a future work.

\vspace{0.5cm}
{\bf Acknowledgement:} The authors thank a lot Prof. Ratbay Myrzakulov for useful suggestions and comments. M.J.S. Houndjo thanks  CNPq/FAPES for financial support.

\end{document}